\documentclass[11pt,onecolumn,amssymb,nofootinbib]{revtex4}
\usepackage{amsmath, amsthm, amscd, amssymb}
\usepackage{bm}
\usepackage{bbm}

\begin{document}

\title{\bf $SU(8)$ family unification with boson--fermion balance}

\author{Stephen L. Adler}
\email{adler@ias.edu} \affiliation{Institute for Advanced Study,
Einstein Drive, Princeton, NJ 08540, USA.}

\begin{abstract}
We formulate an $SU(8)$ family unification model motivated by requiring that
the theory should incorporate the graviton,  gravitinos, and the fermions and gauge fields
of the standard model, with boson--fermion balance.  Gauge field $SU(8)$ anomalies cancel
between the gravitinos and spin $\frac {1}{2}$  fermions. The 56 of scalars breaks $SU(8)$  to $SU(3)_{family} \times SU(5)\times U(1)/Z_5$,
with the fermion representation content needed for ``flipped'' $SU(5)$ with three families, and
with  residual scalars in the $10$ and $\overline{10}$ representations that
break flipped $SU(5)$ to the standard model.  Dynamical symmetry breaking can
account for the generation of $5$ representation scalars needed to break the
electroweak group.  Yukawa couplings of the 56 scalars  to the fermions are forbidden by chiral and gauge symmetries,
so in the first stage of $SU(8)$ breaking fermions remain massless. In the limit of vanishing gauge coupling,
there are $N=1$ and $N=8$ supersymmetries relating the scalars to the fermions, which restrict the form of scalar self-couplings and should improve the
convergence of perturbation theory, if not making the theory finite and ``calculable''.  In an Appendix we give an analysis of symmetry breaking by a Higgs
component, such as the $(1,1)(-15)$ of the $SU(8)$ 56 under $SU(8) \supset SU(3) \times SU(5) \times U(1)$,  which has nonzero $U(1)$ generator.
\end{abstract}

\maketitle

\section{Introduction}
The presence of bosons and fermions in Nature makes the
idea of a fundamental boson--fermion balance appealing, and this has
motivated an extensive search for supersymmetric extensions of the
standard model.  However, since the observed particle
mass spectrum is not supersymmetric,
supersymmetry breaking must be invoked, and despite much effort
a definitive model, and a definitive symmetry breaking mechanism, have
yet to emerge.  We turn in this paper to another possibility, that
boson--fermion balance without full supersymmetry is the relevant property of the
unification theory, and construct a model based on this philosophy
motivated by $SU(8)$ unification and supergravity.

\section{Counting states}
The model we study is inspired by the state structure of maximal $SO(8)$ supergravity.  The usual counting of on-shell states
for $N=8$ supergravity is one graviton with 2 helicity states, 8 Majorana gravitinos with 16 helicity
states, 28 vectors with 56 helicity states, 56 Majorana fermions with 112 helicity states, and 70
scalars with 70 helicity states. Thus there are 2+56+70=128 boson states, and 16+112=128 fermion states, giving
the required boson--fermion balance, and interacting models with this field content exist.  Unfortunately, however,
these models do not contain the full particle and gauge group content needed for the standard model.

In a fascinating comment in his magisterial work on ``Group theory for unified model building'', Richard Slansky wrote \cite{slansky} :
``One may wish to speculate about a future unified theory of all interactions and all elementary particles that would
resemble $SO_8$ supergravity but involve sacrificing some principle now held sacred, so that the notion of extended
supergravity could be generalized.  In such a hypothetical theory, an internal symmetry group $G$ larger than $SO_8$
would be gauged by spin 1 bosons, and both the spin $\frac{3}{2}$ and spin $\frac{1}{2}$ fermions would be assigned
to representations of $G$.  It is then very natural to suppose that the spin $\frac{3}{2}$ fermions would belong
to some basic representation of $G$ and would include only color singlets, triplets and antitriplets.  The spin
$\frac{1}{2}$ particles would then presumably be assigned to a more complicated representation.  These speculations
are a major motivation for this review, as they were for ref. [6].''(Slansky's reference 6 is Gell-Mann, Ramond and Slansky  \cite{gellmann}.)

The rest of this paper proceeds in the spirit of Slansky's remarks (which as we shall see, describe the
model that we construct.)
We begin by noting that if the 70 scalars are eliminated from the counting, and their degrees of freedom are
redistributed to the two helicities of 35 vectors, we are left with 28+35=63 vectors in all, which can
be assigned to the adjoint representation of an $SU(8)$ group.  The remaining representations in the
counting, the 8 and 56, can be interpreted as the fundamental and rank three antisymmetric tensor
representations of $SU(8)$, giving an ``$SU(8)~{\rm graviton}$'' multiplet consisting of the
graviton, the 8 gravitinos, the 63 vectors, and the 56 fermions.  There are still 128 boson
and 128 fermion helicities in this model, but the state structure is no longer the one corresponding
to unitary supersymmetry representations in Hilbert space.  Since we are working in 4 dimensions, and
the model is not supersymmetric, we switch at this point from Majorana fermions to the usual left chiral
(L) Weyl fermions used in grand unification, but the state counting is the same.

There is a long history of $SU(8)$ unification models in the literature; see \cite{curtright}--\cite{martinez}.
Of particular interest  are the papers of Curtright and Freund \cite{curtright}, C. Kim and Roiesnel \cite{kimone}, and J. Kim and Song \cite{kimtwo},
which incorporate spin $\frac{1}{2}$ fermions through single left chiral $\bar 8$, $\overline{28}$, and 56 representations of $ SU(8)$.
Under breaking to $SU(5)$, the $\overline{28}$ of $SU(8)$ contains three copies of the $\bar 5$ of $SU(5)$, and the
$56$ of $SU(8)$ contains three copies of the $10$ of $SU(5)$, so this representation content incorporates the three
standard model families.  Additionally, the paper of Curtright and Freund explicitly ties the representation numbers
8, 28, and 56 to those appearing in $N=8$ supergravity, with the suggestion that the $SU(8)$ gauge bosons may appear
as bound states, as suggested by Cremmer and Julia \cite{cremmer}.

Returning to the ``$SU(8)~{\rm graviton}$'' multiplet, the $56_L$ of fermions contains three families in the $SU(5)$
$10_L$ representation.
In order to incorporate three $SU(5)$ ${\overline 5}_L$ families into a model with boson--fermion balance, we adjoin to the
``$SU(8)~{\rm graviton}$'' multiplet a ``$SU(8)~{\rm matter}$'' multiplet consisting of a complex scalar field
in the $56$ representation of $SU(8)$, and {\it two} copies of a fermion spin $\frac{1}{2}$ field in the
$\overline{28}_L$ representation of $SU(8)$.  Use of a complex scalar is necessary since the $56$ is a complex representation,
and so cannot be assigned to a real scalar multiplet.  Boson--fermion balance then requires that we double
the number of $\overline{28}_L$ representations, so that the number of spin $\frac{1}{2}$ helicity states is $2 \times 2 \times 28 =112$,
equal to the number of helicity states in a complex 56 scalar. (Although boson--fermion balance could be achieved with a
single $28_L$ of fermions and a complex 28 of scalars,  $SU(8)$ anomalies would not cancel, and $SU(8)$ could not be broken to $SU(3)\times SU(5)$.)
The $SU(8)$ fermion and boson content of the model is summarized in Table I.

\begin{table} [h]
\caption{Field content of the model, with the top part of the table showing the ``$SU(8)~{\rm graviton}$'' multiplet, and the bottom
part of the table showing the ``$SU(8)~{\rm matter}$'' multiplet. The linearized graviton  $h_{\mu\nu}$ is defined by $g_{\mu\nu}=\eta_{\mu\nu}+\kappa h_{\mu \nu}$,
with $\eta_{\mu \nu}$ the Minkowski metric and $\kappa$ the gravitational coupling.
Branching rules are from Slansky \cite{slansky} with $U(1)$ generators (or charges) in
parentheses, followed in curly brackets by equivalent $U(1)$ generators modulo 5. (The modulo 5 ambiguities in these assignments have been used to give
the assignments needed for flipped $SU(5)$, plus states that can be paired into condensates or are neutral with
respect to the $SU(3) \times SU(5)\times U(1)/Z_5$ force.)
Square brackets on the field subscripts and superscripts  indicate complete antisymmetrization of the enclosed indices.  The indices $\alpha,\beta,\gamma$
range from 1 to 8,  the index $A$ runs from 1 to 63, and $\mu,\,\nu$ are Lorentz indices.}
\centering
\begin{tabular}{ c c c c c}
\hline\hline
field~~~ & spin~~~ & $SU(8)$ rep.~~~ & helicities~~~ & branching to $SU(3) \times SU(5)\times U(1)$ \\
\hline
$h_{\mu \nu}$ & 2 & 1 & 2&1\\
$\psi_{\mu}^{\alpha}$ & Weyl \,$\frac{3}{2}$ & $8_L$ & 16  & (3,1)(-5)\{0\}+(1,5)(3)\{-2\}\\
$A_{\mu}^A$ & 1 & 63 & 126 & (1,1)(0)\{0\}+(8,1)(0)\{0\}+(3,$\overline{5}$)(-8)\{2\}+($\overline{3}$,5)(8)\{-2\}+(1,24)(0)\{0\}\\
$\chi^{[\alpha\beta\gamma]}$& Weyl \, $\frac{1}{2}$ & $56_L$ & 112 & (1,1)(-15)\{0\}~~~~~~~~~+~~~~~~~~(1,$\overline{10}$)(9)\{-1\}+($\overline{3}$,5)(-7)\{3\}+(3,10)(1)\{1\}\\
\hline
$\lambda_{1\,[\alpha \beta]}$& Weyl\, $\frac{1}{2}$&$\overline{28}_L$ & 56 & (3,1)(10)\{5\}+(1,$\overline{10}$)(-6)\{-1\}+($\overline{3}$,$\overline{5}$)(2)\{-3\}\\
$\lambda_{2\,[\alpha \beta]}$& Weyl \,$\frac{1}{2}$&$\overline{28}_L$ & 56 & (3,1)(10)\{5\}+(1,$\overline{10}$)(-6)\{-1\}+($\overline{3}$,$\overline{5}$)(2)\{-3\}\\
$\phi^{[\alpha\beta\gamma]}$& complex\, 0& 56 & 112 & (1,1)(-15)\{0\}+(1,$\overline{10}$)(9)\{-1\}+($\overline{3}$,5)(-7)\{-2\}+(3,10)(1)\{1\}\\
\hline\hline
\end{tabular}
\label{table:fieldcontent}
\end{table}

\section{Anomaly cancelation}

To have a consistent $SU(8)$ gauge theory, anomalies must cancel.  In the papers of Curtright and Freund \cite{curtright},
Kim and Roiesnel \cite{kimone}, and Kim and Song \cite{kimtwo}, this is achieved through
\begin{align}
{\rm anomaly}(\overline{8}_L) =& -1  ~~~,\cr
{\rm anomaly} (\overline{28}_L)=&-4~~~,\cr
{\rm anomaly} (56_L)=&\,5~~~,\cr
{\rm total ~anomaly}=&-1-4+5=0~~~.
\end{align}

In our model anomaly cancelation involves the same representations, up to conjugation,
but different counting.  Instead of a spin $\frac{1}{2}$ $\overline{8}_L$, our ``$SU(8)~{\rm graviton}$'' multiplet
contains a spin $\frac{3}{2}$ $8_L$.  Since the chiral anomaly of a spin $\frac{3}{2}$ particle is three times that of
the corresponding spin $\frac{1}{2}$ particle \cite{nielsen}, \cite{duff},  the $8_L$ of gravitinos contributes 3 to the anomaly
count.  The $56_L$ of spin $\frac{1}{2}$ fermions contributes 5 as before, while the two $\overline{28}_L$ of spin $\frac{1}{2}$
fermions contribute $-8$, giving
\begin{align}\label{cancel}
3 \times {\rm anomaly}( 8_L) =&\, 3  ~~~,\cr
2 \times {\rm anomaly} (\overline{28}_L)=&-8~~~,\cr
{\rm anomaly} (56_L)=&\,5~~~,\cr
{\rm total ~anomaly~in~our~model}=&\,3-8+5=0~~~.\cr
\end{align}
So anomalies cancel, but by a different
mechanism than in refs. \cite{curtright}, \cite{kimone}, and \cite{kimtwo}.  Anomaly cancellation
with the counting of Eq. \eqref{cancel} (using the conjugate representations $\overline{8}$, 28, and $\overline{56}$ ) was noted
by Marcus \cite{marcus} in a study of dynamical gauging of $SU(8)$ in $N=8$ supergravity.

\section{Gauge symmetry breaking and state content}

We turn to the issue of gauge symmetry breaking.
Symmetry breaking in our model is initiated by a Brout-Englert-Higgs (BEH) mechanism using the
complex scalar field in the ``$SU(8)~{\rm matter}$'' multiplet, which is in the 56 of $SU(8)$.
(This can be accomplished by either an explicit negative mass for the scalar in the action, or by an
alternative that we favor,  the Coleman-Weinberg \cite{colewein} mechanism induced by radiative corrections starting from a massless scalar.)
Since the 56 representation of $SU(8)$ branches to the $56_v$ of $SO(8)$, not to a singlet of $SO(8)$,
the symmetry breaking
pathway of our model cannot pass through $SO(8) \times U(1)$.   Referring to Table I, which gives the branching of the 56 of $SU(8)$  to
$SU(3) \times SU(5) \times U(1)$, we see that there is a singlet (1,1) of $SU(3) \times SU(5)$
with a nonzero $U(1)$ generator of $-15$. Hence there are two interesting symmetry breaking pathways.  In the first, the BEH mechanism breaks  $SU(8)$
to  $SU(3) \times  SU(5)$, with the $U(1)$ gauge symmetry either completely broken or, as discussed in Appendix A, broken to $U(1)/Z$.  It is then natural to identify
the $SU(3)$ factor as a family symmetry group, and the $SU(5)$ factor and fermion content as the usual minimal
grand unification group \cite{georgi}.  In the second, the $U(1)$ gauge symmetry breaks only  to $U(1)/Z_5$, that is, after symmetry
breaking there is an equivalence between values of $U(1)$ generators that differ by multiples of 5, as a result of a periodicity in the $U(1)$ generator of the broken symmetry ground state, which is discussed in detail in Appendix A.
It is again natural to identify the unbroken $SU(3)$ factor as a family symmetry group.  An inspection of the $U(1)$ generators modulo 5, given
in curly brackets in Table I, shows that the fermion content in this breaking pathway contains all the representations needed for flipped $SU(5)$ grand unification \cite{flipped}.

To elaborate on this, the basic flipped $SU(5)$ model \cite{wiki} consists of a $10\{1\}$ for the quark doublet $Q$, the down quark $d^c$, and the right handed neutrino $N$;
a $\overline{5}\{-3\}$ for the lepton doublet $L$ and the up quark $u^c$; and a $1\{5\}$ for the charged lepton $e^c$. Referring to Table I, we see that $\chi$ contains a $(3,10)\{1\}$,
while  $\lambda_2$  contains a $(\overline {3},\overline{5})\{-3\}$ and a $(3,1)\{5\}$.   This gives three $3$ or $\overline{3}$ families of the states needed for basic flipped $SU(5)$.
Note that we have chosen the $U(1)$ charge assignments modulo 5 needed to make this correspondence possible.
This guarantees that the correct particle charge assignments are obtained after
further breaking to the standard model, and also implies that $SU(5)$ anomalies cancel within the set of
spin $\frac{1}{2}$ states assigned to flipped $SU(5)$, without invoking the spin $\frac{3}{2}$ states.  The remaining states are the $(\overline{3},5)\{3\}$ in $\chi$ and
the $(\overline{3},\overline{5})\{-3\}$ in $\lambda_1$, which after family symmetry breaking can pair to form a condensate;  the $(1,1)\{0\}$ in $\chi$,
which does not feel the $SU(3)\times SU(5) \times U(1)/Z_5$ force and could be a dark matter candidate; and the $(3,1)(10)\{5 \equiv 0\}$ in
$\lambda_{1,2}$, which together with the $(\overline{3},5)(-7)\{3\}$ in $\chi$ can form a condensate which leads to the standard model Higgs
through dynamical chiral symmetry breaking (see below).  There are also  three $(1,\overline{10})\{-1\}$, one in each of the fermions $\chi$,
$\lambda_1$, and $\lambda_2$, which after family symmetry $SU(3)$ breaking can form condensates with the $(3,10)\{1\}$ in $\chi$ to affect the particle mass spectrum.
 We note finally that an
extended version of flipped $SU(5)$,  proposed recently by Barr \cite{barr1}, introduces a vector-like pair $5\{-2\equiv 3\} + \overline{5}\{2\equiv -3\}$ in each family and uses them to argue that  proton decay can be rotated away.

There are residual boson states left after the 56 representation boson $\phi$ breaks the group $SU(8)$, with 63 generators, to $SU(3)\times SU(5) $, with
$8+24=32$ generators, plus the single additional generator of the discrete group $U(1)/Z_5$ when $U(1)$ is not completely broken.   Since $63-33=30$ real components of $\phi$ are absorbed to form longitudinal components of the broken $SU(8)$ generators in the $(3,\overline{5})(-8)\{2\}$ and $(\overline{3},5)(8)\{-2\}$ representations, these components can only
come from the  $(\overline{3},5)(-7)\{-2\}$  representation in the branching expansion of Table I.  So the residual boson states necessarily are the representations
$(1,\overline{10})(9)\{-1\}$  and $(3,10)(1)\{1\}$, plus the $(1,1)(-15)\{0\}$ when $U(1)$ is only broken to $U(1)/Z_5$.    Since breaking minimal $SU(5)$ to the standard model requires a scalar in the 24 representation, the
symmetry breaking pathway to $SU(3) \times SU(5)$ with minimal $SU(5)$ requires dynamical generation of this 24, to be further discussed below.

On the other hand, the residual boson states after $SU(8)$ breaking
contain the Higgs boson representations needed to break flipped
$SU(5)$ to the standard
model \cite{flipped}. Elaborating on this,
the basic flipped $SU(5)$ model uses a $\overline{10}\{-1\}$ and a $10\{1\}$ of scalars to
break flipped $SU(5)$ to the standard model, and these representations are
residual components of the scalar $\phi$.
To further  break the electroweak group of the standard model to the
electromagnetic $U(1)$ group, flipped $SU(5)$ requires a $5\{-2\}$ of
scalars, which contains the standard model Higgs.  This is not present
as a residual scalar component of $\phi$, but as shown below can by generated
in our model by dynamical symmetry breaking.

\section{Asymptotic freedom and global symmetries}

The $SU(8)$ representation content
of the model has a small enough spin 0, spin $\frac{1}{2}$, and spin $\frac{3}{2}$  content to keep the theory asymptotically free ,
\begin{align}
&\frac{1}{3}[11 c(1)-26 c({\rm Weyl}\, 3/2)-2 c(\rm {Weyl}\, 1/2)-c({\rm complex}\, 0)] \cr
=&\frac{1}{3}[11\times 16-26 \times 1-2\times(15+2\times 6)-  15]=27>0~~~,\cr
\end{align}
with $c(s)$ the index of the $SU(8)$ representation with spin $s$.   (For the spin $\frac{3}{2}$ beta function see Curtright \cite{curt3}, Duff \cite{duff}, and Fradkin and Tseytlin \cite{fradkin};
the index $c$ is tabulated as $\ell$ in the tables of Slansky \cite{slansky}.)  Thus the
$SU(8)$ coupling increases as the energy decreases, which can trigger dynamical symmetry breaking in addition to the symmetry breaking
 provided by the elementary Higgs fields.
In addition to a locally gauged $SU(8)$ symmetry, our model admits a number of global chiral symmetries associated with the fermion fields \cite{peskin}.  The first is an overall chiral $U(1)$ symmetry associated with an overall $U(1)$ rephasing of all of the fermion fields, spin $\frac{3}{2}$ as well as spin $\frac{1}{2}$.    It will be
convenient to regard this phase as associated with the $8_L$ of spin $\frac{3}{2}$ fermions, labeled $\psi_{\mu}$ in Table I.    We expect this global symmetry to be
broken by the usual instanton and anomaly mechanism that is invoked to solve the ``$U(1)$ problem'' in QCD \cite {thooft}, \cite{coleman}.  The second global symmetry is an overall $U(1)$ rephasing of the spin $\frac{1}{2}$  $56_L$ fermion fields  $\chi$ relative to $\psi_{\mu}$.
Finally, since the kinetic Lagrangian contains the doubled $\overline{28}_L$ representation spanned by the fermion basis $\lambda_{1,2}$, there is a global $U(2)$ symmetry associated with mixing of these basis states, relative to the phase of $\psi_{\mu}$.

\section{Dynamical versus elementary Higgs symmetry breaking}

As already noted, in the $SU(8) \supset SU(3) \times SU(5)$ symmetry breaking pathway, the $SU(5)$ 24 representation needed for breaking to the standard model must
be generated dynamically.  A quick review of the theory of dynamical symmetry breaking is given in  Appendix B. A strategy for getting a 24, following \cite{farhi}, would be to
generate a 24 condensate at the unification scale, which violates the chiral symmetries of the theory, and so leads to a 24 Goldstone boson, which could then
serve as the 24 Higgs.  There are two problems with this scenario.  The first is that the only way to generate a 24 representation of $SU(5)$ from the
representations in Table I is through either $\overline{5} \times 5$ or $\overline{10} \times 10$, both of which contain an $SU(5)$ singlet in addition to
a 24.  Since the singlet is always the most attractive channel \big(see Eq. \eqref{mac}\big), dynamical generation of a 24 seems unlikely \cite{peskin}, \cite{mactheory}, \cite{raby}.  The second problem is that
if gauge couplings were strong enough for a 24 condensate to be formed at the unification scale, then one would expect  that unification scale condensates involving the wanted fermions in the $(3,10)$ representation in Table I would also form, removing these states from the low energy spectrum.  So getting the standard model from our theory  through an
$SU(8) \supset SU(3) \times SU(5)$ symmetry breaking pathway is not plausible.

The situation is more favorable for the  $SU(8) \supset SU(3) \times SU(5) \times U(1)/Z_5$ symmetry breaking pathway, which as explained in Appendix A involves a  ground state that is periodic in the $U(1)$ generator.
This pathway does not require dynamical condensates  to break $SU(3) \times SU(5)\times U(1)/Z_5$ to   the standard model; the residual elementary scalar states in $\phi$ can do this, as well as breaking  the $SU(3)$ family symmetry. (Before $SU(5)$ breaking, the $(3,10)(1)\{1\}$ scalar can break family
$SU(3)$ to $SU(2)\times U(1)$, accommodating two light families and one heavy one, while after $SU(5)$ breaking family symmetry
can be completely broken.\cite{li}) However, we have seen that the residual components of the 56 scalar, after $SU(8)$ symmetry breaking, do not contain the
$5\{-2\}$ needed in flipped $SU(5)$ to break the electroweak symmetry of the standard model, so here the dynamical symmetry breaking mechanism of
Ref. 29 is needed.  Referring to Table 1, we see that the $(\overline{3},5)(-7)$ component of $\chi$ can
pair with the $(3,1)(10)$ component of the doublet $\lambda_a$, to form a doublet condensate which is in the representation $(1,5)(3)\{3\equiv-2\}$, with the
family $SU(3)$ group acting as the hypercolor or ``technicolor'' force in binding the condensate in the most attractive
singlet channel.  Since this condensate breaks the $U(2)\big(=U(1)\times SU(2)\,\big)\times U(1)$ global chiral symmetry of the doublet $\lambda_a$ and of $\chi$ to a diagonal $U(1)$, there will be a $U(1)$ singlet and a $SU(2)$ triplet of Goldstone bosons with the needed flipped $SU(5)$ quantum numbers  $5\{-2\}$.
These Goldstone bosons are still gauged under the  $SU(5)$  group, so the Coleman-Weinberg mechanism will generate
symmetry breaking potentials for them, leading to the electroweak symmetry breaking of the standard model. Our model thus suggests
that in addition to the observed (presumably singlet) Higgs boson, there should also be an $SU(2)$ triplet of Higgs bosons with the
same standard model quantum numbers.

\section{The gauge sector action}

We turn now to writing down the action for the gauge sector of our model.  Since all fermion representations are antisymmetrized direct
products of fundamental 8 representations, we need only use generators $t_A$ for the fundamental 8 of $SU(8)$ to construct
covariant derivatives of the fermion fields.  We follow here the conventions of \cite{freedman}, and take the $t_A$ to be anti-self-adjoint,
with commutators and trace normalization given by
\begin{align}
[t_A,t_B]=&f_{ABC}t_C~~~,\cr
{\rm Tr}(t_A t_B)= -\frac{1}{2}&\delta_{AB}~~~,\cr
\end{align}
with implicit summation on repeated indices.

Defining the gauge variation of the gauge potential by
\begin{equation}
\delta_G A_{\mu}^A=\frac{1}{g} \partial_{\mu}\Theta^A + f_{ABC}A_{\mu}^B \Theta^C~~~,
\end{equation}
the gauge covariant field strength $F_{\mu \nu}^A$ is defined as
\begin{equation}
F_{\mu \nu}^A=\partial_{\mu}A_{\nu}^A-\partial_{\nu}A_{\mu}^A + g f_{ABC} A_{\mu}^B A_{\nu}^C~~~,
\end{equation}
and has the gauge variation
\begin{equation}
\delta_G F_{\mu \nu}^A=f_{ABC}F_{\mu \nu}^B \Theta^C~~~.
\end{equation}

We can now define covariant derivatives of the fermion fields of the model.  Writing
\begin{equation}\label{adef}
A_{\mu \, \beta}^{\alpha}=A_{\mu}^A (t_A)^{\alpha}_{~~\beta}~~~,
\end{equation}
the covariant derivatives of the fermion fields $\psi_{\mu}^{\alpha}$,  $\chi^{[\alpha\beta\gamma]}$ and
$\lambda_{a\,[\alpha \beta]}$, ~$a=1,2$ of
Table I (in the 8, 56, and $\overline{28}$  representations respectively) are defined by
\begin{align}
D_{\nu} \psi_{\mu}^{\alpha}=& \partial_{\nu} \psi_{\mu}^{\alpha}+g A_{\nu \, \delta}^{\alpha} \psi_{\mu}^{\delta}~~~,\cr
D_{\nu} \chi^{[\alpha\beta\gamma]}=& \partial_{\nu}  \chi^{[\alpha\beta\gamma]}+ g(A_{\nu \, \delta}^{\alpha}  \chi^{[\delta\beta\gamma]}
+ A_{\nu \, \delta}^{\beta}\chi^{[\alpha\delta\gamma]}+A_{\nu \, \delta}^{\gamma}\chi^{[\alpha\beta\delta]})~~~,\cr
D_{\nu} \lambda_{a\,[\alpha\beta]}=&\partial_{\nu} \lambda_{a[\alpha\beta]}+g(A_{\nu \, \alpha}^{\delta}\lambda_{a[\delta\beta]}+
A_{\nu \, \beta}^{\delta}\lambda_{a[\alpha\delta]})~~~,~a=1,2~~~.\cr
\end{align}
Similarly, for the scalar field $\phi^{[\alpha\beta\gamma]}$, the covariant derivative is defined by
\begin{equation}
D_{\nu} \phi^{[\alpha\beta\gamma]}= \partial_{\nu}  \phi^{[\alpha\beta\gamma]}+ g(A_{\nu \, \delta}^{\alpha}  \phi^{[\delta\beta\gamma]}
+ A_{\nu \, \delta}^{\beta}\phi^{[\alpha\delta\gamma]}+A_{\nu \, \delta}^{\gamma}\phi^{[\alpha\beta\delta]})~~~.
\end{equation}
These give
\begin{equation}
\delta_G \psi_{\mu}^{\alpha}=-\theta^A t_{A\delta}^{\alpha} \psi_{\mu}^{\delta}~~,~~~
\delta_G D_{\nu}\psi_{\mu}^{\alpha}=-\theta^A t_{A\delta}^{\alpha} D_{\nu}\psi_{\mu}^{\delta}~~~,
\end{equation}
and similarly for the gauge variations of the other fields and their covariant derivatives.

With the $SU(8)$ covariant derivatives of the fields defined, we can now write down the gauge sector action of the model,
with gravity treated in the linearized approximation, as follows. The total action is
\begin{align}
S({\rm total})=&S(h_{\mu\nu})+S(\psi_{\mu})+S(A_{\mu})+S(\chi)+S(\lambda_{1,2})\cr
+&S_{\rm kinetic}(\phi)+S_{\rm self-coupling}(\phi)+S_{\rm fermion-coupling}(\phi,\psi_{\mu},\chi,\lambda)~~~.\cr
\end{align}
For $S(h_{\mu\nu})$ we have the usual linearized gravitational action,
\begin{align}
S(h_{\mu\nu})=&\frac{1}{8} \int d^4x  h^ {\mu\nu} H_{\mu\nu}~~~,\cr
H_{\mu\nu}=& \partial_{\mu}\partial_{\nu} h_{\lambda}^{\lambda}+\Box h_{\mu\nu}-\partial_{\mu}\partial^{\lambda} h_{\lambda \nu}
-\partial_{\nu}\partial^{\lambda} h_{\lambda \mu}-\eta_{\mu\nu} \Box h_{\lambda}^{\lambda} + \eta_{\mu \nu} \partial^{\lambda}\partial^{\rho} h_{\lambda\rho}~~~.\cr
\end{align}
For the gravitino action we have the $SU(8)$ gauged extension of the usual expression,
\begin{align}
S(\psi_{\mu})=&\frac{1}{2} \int d^4x \overline{\psi}_{\mu\,\alpha} R^{\mu\,\alpha} ~~~,\cr
R^{\mu\,\alpha}=& i \epsilon^{\mu\eta\nu\rho} \gamma_5 \gamma_{\eta} D_{\nu} \psi_{\rho}^{\alpha}=R^{\mu\,\alpha}_{\rm free}+R^{\mu\,\alpha}_{\rm interaction}~~~,\cr
R^{\mu\,\alpha}_{\rm free}=&i \epsilon^{\mu\eta\nu\rho} \gamma_5 \gamma_{\eta} \partial_{\nu} \psi_{\rho}^{\alpha}~,~~
R^{\mu\,\alpha}_{\rm interaction}=i \epsilon^{\mu\eta\nu\rho} \gamma_5 \gamma_{\eta} A^{\alpha}_{\nu\delta} \psi_{\rho}^{\delta}~~~.\cr
\end{align}
Since the free gravitino action is invariant under the gravitino  gauge transformation $\psi_{\rho}^{\alpha} \to \psi_{\rho}^{\alpha}+\partial_{\rho}\epsilon^{\alpha}$, a gauge
fixing condition is needed to quantize, which can be taken in the covariant form $\gamma^{\rho}\psi_{\rho}^{\alpha}=0.$  The associated ghost fields then play a role
in the spin $\frac{3}{2}$ anomaly calculation \cite{alvarez}.

The $SU(8)$ gauge field action has the standard form
\begin{equation}
S(A_{\mu})=-\frac{1}{4}\int d^4x  F^A_{\mu\nu}F^{A\mu\nu}~~~,
\end{equation}
and the spin $\frac{1}{2}$  fermion actions are
\begin{align}
S(\chi)=&-\frac{1}{2}\int d^4x  \overline{\chi}_{[\alpha\beta\gamma]}\gamma^{\nu}D_{\nu}\chi^{[\alpha\beta\gamma]}~~~,\cr
S(\lambda_{1,2})=&- \frac{1}{2}\int d^4x \sum_a \overline{\lambda_a}^{[\alpha\beta]} \gamma^{\nu}D_{\nu} \lambda_{a\,[\alpha\beta]}~~~,\cr
\end{align}
where we have written the second line in a form which exhibits its global $U(2)$ invariance.
Finally, for the scalar field kinetic action we have
\begin{equation}
S_{\rm kinetic}(\phi)=-\frac{1}{2}\int d^4x   (D^{\nu} \phi)^*_{[\alpha\beta\gamma]}D_{\nu} \phi^{[\alpha\beta\gamma]} ~~~.
\end{equation}

For later use, we note that the equations of motion of the spin $\frac{1}{2}$ fermions and the spin 0 boson, following from these gauged kinetic actions but ignoring
for the moment possible additional scalar interaction terms, are
\begin{align}\label{eqmo}
\gamma^{\nu}D_{\nu}\chi^{[\alpha\beta\gamma]}=&0~~~,\cr
\gamma^{\nu}D_{\nu} \lambda_{a\,[\alpha\beta]}=&0~,~~~a=1,2~~~,\cr
D^{\nu}D_{\nu} \phi^{[\alpha\beta\gamma]}=&0~~~.\cr
\end{align}

\section{Absence of scalar--fermion Yukawa couplings}

We turn next to possible Yukawa couplings $S_{\rm fermion-coupling}(\phi,\psi_{\mu},\chi,\lambda)$, which we show must all vanish.
The chirality requirements for forming nonzero Yukawa couplings are the same as those for forming condensates discussed in Appendix B.
Thus, chirality requires that Yukawa couplings of the spin $\frac{1}{2}$ fermions must be of the form $\Psi_{L1}^T i \gamma^0 \Psi_{L2} \Phi$,
with $\Psi_{1,2}$ any of the spin $\frac{1}{2}$ fermion fields, and $\Phi$ either $\phi$ or $\phi^*$, with $SU(8)$ indices contracted
to form a singlet.  But this is not possible, since the product of two $\chi$ has 6 upper $SU(8)$ indices, the product of two $\lambda$ has
4 lower $SU(8)$ indices, and the product of $\chi$ with a $\lambda$ has 3 upper $SU(8)$ indices and two lower $SU(8)$ indices, none of
which can be contracted with a $\phi$, with three upper indices, or a $\phi^*$, with three lower indices, to form an $SU(8)$ singlet.
Hence there are no Yukawa couplings involving the spin $\frac{1}{2}$ fermions by themselves.  This implies that after $SU(8)$ symmetry breaking, the spin $\frac{1}{2}$ fermions remain massless, which is essentialfor getting a three family flipped $SU(5)$ model.

Yukawa couplings of a spin $\frac{3}{2}$ field to a spin $\frac{3}{2}$ field and the scalar are forbidden by a chirality and $SU(8)$
index contraction argument similar to that used in the case of two spin $\frac{1}{2}$ fields.  This argument does not forbid couplings of
a spin $\frac{3}{2}$ field to a spin $\frac{1}{2}$ field and the scalar of the form $\overline{\lambda_a}^{[\alpha\beta]} \gamma^{\nu} \psi_{\nu}^{\gamma} \phi^*_{[\alpha \beta \gamma]}$
and its conjugate, but these vanish when the gravitino gauge fixing condition $ \gamma^{\nu} \psi_{\nu}^{\gamma} =0$ is imposed.

\section{Supersymmetries in the limit of zero gauge coupling}

Let us now consider the free limit of the theory in which the gauge coupling $g$ vanishes, so that the covariant derivatives $D_{\nu}$ become
ordinary partial derivatives $\partial_{\nu}$, and the equations of motion of Eq. \eqref{eqmo} simplify to
\begin{align}\label{eqmo1}
\gamma^{\nu}\partial_{\nu}\chi^{[\alpha\beta\gamma]}=&0~~~,\cr
\gamma^{\nu}\partial_{\nu} \lambda_{a\,[\alpha\beta]}=&0~,~~~a=1,2~~~,\cr
\partial^{\nu}\partial_{\nu} \phi^{[\alpha\beta\gamma]}=&0~~~.\cr
\end{align}
One can then form two conserved $SU(8)$  representation 8 supercurrents,
\begin{align}\label{8current}
J^{\mu\,\alpha}_{a}=&\gamma^{\nu}(\partial_{\nu}\phi^{[\alpha\beta\gamma]})\gamma^{\mu}\lambda_{a[\beta\gamma]}~~~,\cr
\partial_{\mu}J^{\mu\,\alpha}_{a}=&0~~~,~~\alpha=1,...,8 ~{\rm and}~ a=1,2~~~,\cr
\end{align}
and an $SU(8)$ singlet conserved supercurrent,
\begin{align}\label{1current}
J^{\mu}=&\gamma^{\nu}(\partial_{\nu}\phi^*_{[\alpha\beta\gamma]})\gamma^{\mu}\chi^{[\alpha\beta\gamma]}~~~,\cr
\partial_{\mu}J^{\mu}=&0~~~.\cr
\end{align}
In deriving supercurrent conservation we have used the equations of motion together with
\begin{equation}
\gamma^{\nu}\gamma^{\mu} \partial_{\nu}\partial_{\mu} \Phi=\eta^{\mu\nu}\partial_{\nu}\partial_{\mu} \Phi~~~,
\end{equation}
which is a consequence of the commutativity of partial derivatives.
The invariance transformation of the free action for which $J^{\mu\,\alpha}_{a}$ (with $a=$  1 or 2) is the
Noether current is
\begin{align}\label{8trans}
\delta \phi^{[\alpha\beta\gamma]}=&\overline{\lambda}_a^{[[\beta \gamma]} \epsilon^{\alpha]}~~~,\cr
\delta \phi^*_{[\alpha\beta\gamma]}=&\overline{\epsilon}_{[\alpha}\lambda_{a[\beta\gamma]]}~~~,\cr
\delta \lambda_{a[\alpha\beta]}=&\gamma^{\nu}\partial_{\nu} \phi^*_{[\alpha\beta\delta]} \epsilon^{\delta}~~~,\cr
\delta \overline{\lambda}_a^{[\alpha\beta]}=&-\overline{\epsilon}_{\delta}\gamma^{\nu}\partial_{\nu} \phi^{[\alpha\beta\delta]}~~~,\cr
\end{align}
and the transformation for which $J^{\mu}$ is the Noether current is
\begin{align}\label{1trans}
\delta \phi^{[\alpha\beta\gamma]}=&\overline{\epsilon} \chi^{[\alpha\beta\gamma]}~~~,\cr
\delta \phi^*_{[\alpha\beta\gamma]}=&\overline{\chi}_{[\alpha\beta\gamma]} \epsilon~~~,\cr
\delta \chi^{[\alpha\beta\gamma]}=&\gamma^{\nu}\partial_{\nu} \phi^{[\alpha\beta\gamma]}\epsilon~~~,\cr
\delta \overline{\chi}_{[\alpha\beta\delta]}=&-\overline{\epsilon} \gamma^{\nu}\partial_{\nu} \phi^*_{[\alpha\beta\gamma]}~~~.\cr
\end{align}

Since covariant derivatives do not commute, when gauge interactions are included there are no longer conserved supercurrents.
For example, if we redefine the singlet current as
\begin{equation}
J^{\mu}=\gamma^{\nu}(D_{\nu}\phi^*_{[\alpha\beta\gamma]})\gamma^{\mu}\chi^{[\alpha\beta\gamma]}~~~,
\end{equation}
then we find
\begin{align}
\partial_{\mu}J^{\mu}=&(D_{\mu} \gamma^{\nu}(D_{\nu}\phi^*_{[\alpha\beta\gamma]}))\gamma^{\mu}\chi^{[\alpha\beta\gamma]}
+ \gamma^{\nu}(D_{\nu}\phi^*_{[\alpha\beta\gamma]})\gamma^{\mu} D_{\mu}\chi^{[\alpha\beta\gamma]} \cr
=&\frac{1}{2}([D_{\mu},D_{\nu}] \gamma^{\nu\mu} \phi^*_{[\alpha\beta\gamma]})\chi^{[\alpha\beta\gamma]}\cr
=&\frac{1}{2}g \gamma^{\nu\mu}(F_{\mu\nu\, \alpha}^\delta \phi^*_{[\delta\beta\gamma]}+F_{\mu\nu\, \beta}^\delta \phi^*_{[\alpha\delta\gamma]}+
F_{\mu\nu\, \gamma}^\delta \phi^*_{[\alpha\beta\delta]})\chi^{[\alpha\beta\gamma]}~~~,\cr
\end{align}
with $\gamma^{\nu\mu}=\frac{1}{2}[\gamma^{\nu},\gamma^{\mu}]$.

\section{Scalar sector self-couplings}

We consider finally the action terms involving the scalar field without gauging.  For the scalar field self-coupling action, taking index permutation possibilities into account, we have
\begin{equation}\label{scalarcoupl}
S_{\rm self-coupling}(\phi)=
\phi^*_{[\rho\kappa\tau]} \phi^*_{[\alpha\beta\gamma]}(g_1  \phi^{[\rho\kappa\tau]}  \phi^{[\alpha\beta\gamma]} +g_2  \phi^{[\alpha\kappa\tau]} \phi^{[\rho\beta\gamma]})~~~,
\end{equation}
which is a straightforward generalization of the usual real scalar field $\phi^4$ coupling.  However, when the gauge coupling $g$ is zero, the kinetic action is invariant under
the supersymmetry transformations of Eqs. \eqref{8trans}   and  \eqref{1trans}, which are not invariances of the self-coupling action of Eq. \eqref{scalarcoupl}.  Hence the couplings $g_1$ and $g_2$ must
be of order $g^2$ or higher order in the gauge coupling.  An important question to be answered is how the invariances of the action affect the renormalization of
$g_{1,2}$: In what order of $g^2$ do they contain logarithms of the ultraviolet cutoff, or are they finite and calculable to all orders?  Since there are no Yukawa couplings,
it is possible that the theory is calculable in the sense suggested by Weinberg \cite{swein}.

\section{Discussion}

Grand unification has been intensively investigated for over forty years, and many different
approaches have been tried.  The model proposed here involves three ingredients that do not
appear in the usual constructions:  (1) boson--fermion balance without full supersymmetry,
(2) canceling the spin $\frac{1}{2}$ fermion gauge anomalies against the anomaly from
a gauged spin $\frac{3}{2}$ gravitino, and (3) using a scalar field
representation  with non-zero $U(1)$ generator to break the gauge symmetry, through a
ground state with periodic $U(1)$ generator structure.  The model has a number of promising
features:  (1) natural incorporation of three families, (2) incorporation of the experimentally
viable flipped $SU(5)$ model, (3) a symmetry breaking pathway to the standard model using the
scalar field required by boson--fermion balance,  together with a stage of most attractive channel
dynamical symmetry breaking, without postulating
additional Higgs fields, and (4) vanishing of bare Yukawa couplings
and zero gauge coupling supersymmetries, which keeps the spin $\frac{1}{2}$ fermions massless
after $SU(8)$ symmetry breaking, and  may improve the
predictive power of the theory.

This investigation started from an attempt to base a supersymmetric theory on the state counting of Sec. II.  In the free limit
of zero $SU(8)$ couplings, we saw that the supercurrents of Eqs. \eqref{8current} and \eqref{1current} are conserved, but that
the analogous construction does not give a conserved supercurrent when  $SU(8)$ gauge interactions are included. Moreover, even in the free limit, there
is no corresponding conserved representation 8 supercurrent for the ``$SU(8)~{\rm gravity}$'' multiplet, since in $SU(8)$, $8\times 63$ does not
contain the totally antisymmetric 56 representation.  If one instead looks for an $SO(8)$ representation 8 supercurrent, a similar problem
arises, since the direct product of 8 with the symmetric 35 of $SO(8)$ again does not contain the totally antisymmetric
56 representation.  So for these reasons we abandoned the search for a supersymmetric model, and instead turned to the
weaker condition of boson--fermion balance.  (Group representation considerations leave open the possibility of constructing
8 $N=1$ Lorentz and gauge non-covariant supercurrents in the free limit, by stacking the two helicity components of the symmetric 35 of the gauge field into an artificial 70 component``scalar'' $\tilde{\phi}^{[\alpha \beta \gamma \delta]}$.)

Many open issues remain.  In Table I, we used the modulo 5 freedom of the $U(1)/Z_5$ charges to assign these charges so that the
representations needed for flipped $SU(5)$ have the usual $U(1)$  charge assignments for that model. This recipe is {\it ad hoc},
and needs further justification from a detailed study of the dynamics of symmetry breaking with a modular ground state prior to dynamical
symmetry breaking.   (For example, it would suffice to show that after dynamical symmetry breaking, charge states differing from the wanted ones are separated by
a large mass gap, or are absent from the asymptotic spectrum altogether,  from anomaly considerations  and/or a discrete analog of the familiar
vacuum alignment condition \cite{vacuum}.)
As is clear, our analysis is focussed solely on boson--fermion balance, Lorentz structures, and group theory, and does not address
further dynamical issues such as running couplings,  proton decay, generating the standard model mass and mixing parameters, $CP$ violation, and flavor changing
neutral current constraints on a multiple Higgs structure.     Nonetheless, the issues examined are
an essential first step in trying to set up a realistic unification model, and the results look promising; the pieces appear to fit together
in a jigsaw puzzle-like fashion reminiscent of what one finds in the standard model.

The model presented here should have distinctive experimental signatures.  First, in common with generic flipped $SU(5)$  models, it will have
three families of sterile neutrinos.  Second, as noted above, in addition to a singlet Higgs boson, it should have an $SU(2)$ triplet of
Higgs bosons with the same quantum numbers.

If the model we propose turns out to be the
path that Nature follows, there will remain  the further question of how the  ``$SU(8)~{\rm gravity}$'' multiplet and the
 ``$SU(8)~{\rm matter}$'' multiplet of the model are unified in a more fundamental structure, for example, as arising from involutions
of a large finite group or from periodic or aperiodic tilings of a large lattice.  We note that the ``$SU(8)~{\rm gravity}$'' multiplet
has 128 boson and fermion helicity states, and the  ``$SU(8)~{\rm matter}$'' multiplet  has 112 boson and fermion helicity states.
These numbers respectively match the numbers of half-integer and integer roots of the exceptional group $E(8)$.  Is this a numerical
coincidence, or a hint of a deep connection with the $E(8)$ root lattice?

\section{Acknowledgements}

I wish to acknowledge the hospitality during the summers of 2013 and 2014 of the Aspen Center for Physics, which is supported by the
National Science Foundation under Grant No. PHYS-1066293. I wish to thank Edward Witten for instructive comments on the
$Z_5$ and $Z_6$ factors in  the Wikipedia article on flipped $SU(5)$, Paul Langacker for reading through the paper and a helpful email correspondence,  Michael Peskin for an informative survey lecture on
composite Higgs models at the Princeton Center for Theoretical Science, and Michael Duff for email correspondence about the spin $\frac{3}{2}$ anomaly and beta function
values, which brought \cite{duff} and \cite{marcus} to my attention.   I also wish to thank Graham Kribs for
giving me the opportunity to talk at the Aspen workshop that he co-organized, and David Curtin for asking a question
during my talk about the masses of the scalars.

\appendix
\numberwithin{equation}{section}

\section{Higgs mechanism using a representation with nonzero $U(1)$ charge}
In the usual application of the Higgs mechanism to grand unification, such  as in the breaking of
minimal $SU(5)$ to the standard model $SU(3) \times SU(2) \times U(1)_Y$, a Higgs representation
is chosen which contains a component that is a singlet under all three factors of the standard model
symmetry group.  Thus, the 24 of SU(5) can be used, since it branches according to
$24=(1,1)(0)+(3,1)(0)+(2,3)(-5)+(2,\overline{3})(5)+(1,8)(0)$, which contains the overall
singlet $(1,1)(0)$.   This singlet can attain a nonzero expectation in a ground state (the ``vacuum'')
that has a definite value 0 of the unbroken $U(1)$ generator.

In the $SU(8)$ model studied in this paper, only the 56 representation is available as a scalar
to break the symmetry to $SU(3)\times SU(5)\times U(1)$, and the component $\phi_{(1,1)(-15)}$ that is an $SU(3)
\times SU(5)$ singlet has nonzero $U(1)$ charge $-15$.  By the generalized Wigner-Eckart
theorem, this component cannot acquire a nonzero expectation in a ground state $|\Omega \rangle$ that is
a $U(1)$ eigenstate with a definite generator value. To get a nonzero expectation, we must take $|\Omega\rangle$
to be a superposition of at least two $U(1)$ eigenstates that differ in their $U(1)$ generators by 15.
Anticipating that we want the final result to have a modulo 5 (and not a modulo 15 or modulo 3) structure, we
write the ground state as a superposition of $U(1)$ eigenstates displaced from one another by 5. Let
$G$ be the $U(1)$ generator, and $|n\rangle$ a $SU(3) \times SU(5)$ singlet that is a  $U(1)$ eigenstate
with eigenvalue \big(or $U(1)$ charge\big) $n$, so that $G|n\rangle = n |n\rangle$.  Then we write the ground state $|\Omega\rangle$
in the form
\begin{equation}
|\Omega\rangle= \sum_n f(n)|5n\rangle ~~~,
\end{equation}
which for generic $f(n)$ completely breaks the $U(1)$ invariance,
\begin{equation}
\langle \Omega | \phi_{(1,1)(-15)} |\Omega \rangle \neq 0~~~.
\end{equation}

As in the similar
analysis of the ground state structure of quantum chromodynamics, let us now impose the requirement of
clustering.  In order for the ground state of a tensor product composite system
\begin{equation}
|\Omega_{A+B}\rangle= \sum_{n_A,n_B} f(n_A+n_B)|A;5n_A\rangle|B;5n_B\rangle
\end{equation}
to factor when the subsystems $A,\,B$ are widely separated,
\begin{align}
|\Omega_{A+B}\rangle=&|\Omega_A\rangle |\Omega_B\rangle~~~,\cr
|\Omega_A\rangle=& \sum_{n_A} f(n_A) |A;5n_A\rangle~~~, \cr
|\Omega_B\rangle=& \sum_{n_B} f(n_B) |B;5n_B\rangle~~~, \cr
\end{align}
we must require $f(n)$ to obey
\begin{equation}\label{fform}
f(n_A+n_B)=f(n_A) f(n_B)~~~.
\end{equation}
This requires
that $f(n)$ must have the functional form
\begin{equation}
f(n)=e^{nz}
\end{equation}
for some complex number $z$. Boundedness as $|n|\to \infty$ requires that $|e^{z}|=1$, so $e^z$ is a phase $e^{i\omega}$.  The ground state then has the form
\begin{equation}
|\Omega\rangle= \sum_{n=-\infty}^{\infty} e^{in\omega}|5n\rangle ~~~,
\end{equation}
and $U(1)$  charges are only conserved modulo 5.  This
ground state corresponds to breaking $SU(8)$ to
$SU(3) \times SU(5) \times U(1) / Z_5$, which is the ``second symmetry breaking pathway'' and the one
chosen for our analysis.  (The ``first symmetry breaking pathway'' corresponds either to breaking $U(1)$
completely by using a non-exponential $f(n)$ that violates clustering, or to breaking $U(1)$ to $U(1)/Z$ by
choosing the ground state $|\Omega\rangle=\sum_{n=-\infty}^{\infty} \exp{(in\omega)} |n\rangle$, which equivalences the integer
$U(1)$ charges all to zero.)

The full basis of states for the second pathway has the form
\begin{equation}
|k\rangle=\sum_{n=-\infty}^{\infty} e^{in\omega}|5n+k\rangle ~~~,
\end{equation}
with $|k=0 \rangle=|\Omega\rangle$.  Under a modulo 5 shift we have
\begin{equation}\label{shift}
|k+5s\rangle=\sum_{n=-\infty}^{\infty} e^{in\omega}|5n+k+5s\rangle
=e^{-is \omega}\sum_{n=-\infty}^{\infty} e^{in\omega}|5n+k\rangle = e^{-is\omega} |k\rangle~~~,
\end{equation}
and so the the state basis has a modulo 5 structure up to overall phases.  Denoting by $G_{\pm}$
the raising and lowering operators on the original basis states $|n\rangle$,
\begin{equation}
G_{+}|n\rangle=|n+1\rangle~,~~~G_{-}|n\rangle=|n-1\rangle~~~,
\end{equation}
we can rewrite Eq. \eqref{shift} as
\begin{equation}
G_{+}^{5s} |k\rangle=e^{-is\omega} |k\rangle~,~~G_{-}^{5s} |k\rangle=e^{is\omega} |k\rangle~~~.
\end{equation}
Using the generalized Wigner-Eckart theorem, we can relate the ground state expectation of
$\phi_{(1,1)(-15)}$ to a constant $K$ times the expectation of $G_{-}^{15}$,
\begin{equation}
\langle \Omega|\phi_{(1,1)(-15)}|\Omega \rangle=K \langle \Omega|G_{-}^{15}|\Omega \rangle
=K e^{3i\omega} \langle \Omega|\Omega \rangle \neq 0~~~.
\end{equation}
So within the modulo 5 state structure, $\phi_{(1,1)(-15)}$ can attain a nonzero ground
state expectation.

\section{Review of condensate formation}

We first review the Lorentz kinematics of forming condensates from Dirac spinors, and then turn to
the dynamics of condensate formation.  For any two Dirac spinors $\Psi_1$ and $\Psi_2$, both $\overline{\Psi_1} \Psi_2$
and $\overline{\Psi_1^c} \Psi_2$ are Lorentz scalars, with $c$ denoting charge conjugation and with $\overline{\Psi}=\Psi^{\dagger} i \gamma^0$.
In analyzing condensate formation, it is
convenient to use real Majorana representation $\gamma_{\mu}$ matrices, with $\gamma_5$ self-adjoint and
skew symmetric, and $\gamma^0$ skew symmetric.  The chiral projectors  $P_L,\,P_R$ defined by
\begin{equation}
P_L=\frac{1}{2}(1+\gamma_5)~~,~~P_R=\frac{1}{2}(1-\gamma_5) ~~~,
\end{equation}
then obey
\begin{align}\label{properties}
P_L^{\dagger}=&P_L\,,~P_R^{\dagger}=P_R~~~,\cr
P_L^T= &P_R\,,~P_R^T=P_L~~~,\cr
\end{align}
with $\dagger$ the adjoint and $T$ the Dirac transpose.
Charge conjugation now reduces to complex conjugation, and so we have
\begin{align}\label{chargeconjg}
\Psi^c=&\Psi^*~~~,\cr
\overline{\Psi^c}= &(\Psi^c)^{\dagger} i\gamma^0 = \Psi^T i \gamma^0~~~.\cr
\end{align}
For left chiral spinors, $\overline{\Psi_{L1}} \Psi_{L2}=0$, while Eqs. \eqref{properties} and \eqref{chargeconjg}
imply that
\begin{equation}\label{condens}
\overline{\Psi_{L1}^c} \Psi_{L2}=\Psi_{L1}^T i \gamma^0 \Psi_{L2} \neq 0~~~.
\end{equation}
Thus Eq. \eqref{condens} gives the general Lorentz structure of scalar
condensates constructed from left chiral spinors.  Since $\gamma_0$ is skew symmetric, and since spinors anticommute,
Eq. \eqref{condens} has the same form when state labels $1,\,2$ are interchanged,
\begin{equation}
\Psi_{L1}^T i \gamma^0 \Psi_{L2}=\Psi_{L2}^T i \gamma^0 \Psi_{L1}~~~.
\end{equation}
Because  this equation involves no complex conjugation, the group representation
content of the condensate is simply the direct product of the representation
content of $\Psi_1$ and $\Psi_2$.

The only way to rigorously determine if condensates form in a theory is to calculate the effective action \cite{cornwall} governing
condensate formation, and this is generally not feasible. So to study the dynamics of condensate formation, one falls back on simple rules of thumb, such as
determining whether the leading order perturbation theory force between the constituents is attractive. The single gluon exchange potential \cite{raby}, \cite{peskin} produced when
a vector gluon mediates the reaction $A+B \to A+B$ is
\begin{align}\label{mac}
V=&\frac{g^2K(A+B;A,B)}{2r}~~~,\cr
K(A+B;A,B)=&C_2(A+B)-C_2(A)-C_2(B)~~~,\cr
\end{align}
with $g$ the gauge coupling and the $C_2$ the relevant Casimirs. (The Casimir for a representation $R$ is calculated from the index $\ell(R)$, the dimension $N(R)$, and the
dimension of the adjoint representation $N({\rm adjoint})$ by $C_2(R)= \ell(R) N({\rm adjoint})/N(R)$; see Slansky \cite{slansky}.) When more than one non-Abelian group acts on
the fermions forming the condensate, the one gluon exchange potentials associated with each are added.


\begin{thebibliography}{99}

\bibitem{slansky}  R. Slansky, Phys. Reports  79, 1 (1981).

\bibitem{gellmann}  M. Gell-Mann, P. Ramond, and R. Slansky, Rev. Mod. Phys. 50, 721 (1978).

\bibitem{curtright}  T. L. Curtright and P. G. O. Freund, ``$SU(8)$ Unification and
Supergravity'', in {\it Supergravity}, P. van Nieuwenhuizen and D. Z. Freedman eds.,
North-Holland (1979).

\bibitem{ramond}  P. Ramond, ``The Family Group  in Grand Unified Theories'', invited talk at the
Sannibel Symposia, Feb. 1979, also arXiv:hep-ph/9809459.


\bibitem{frampton} P. H. Frampton, Phys. Lett. B 89, 352 (1980).

\bibitem{chakrabarti}  J. Chakrabarti, M Popovi\'c, and R. N. Mohapatra, Phys. Rev.
D 21, 3212 (1980).

\bibitem{kimone}  C. W. Kim and C. Roiesnel, Phys. Lett. B 93, 343 (1980).

\bibitem{chkareuli1} Dzh. L. Chkareuli, Pis'ma Zh. Eksp. Teor. Fiz. 32, 684 (1980).

\bibitem{kimtwo}  J. E. Kim and H. S. Song, Phys. Rev. D 25, 2996 (1982).

\bibitem{yun1} S. K. Yun, Phys. Rev. D 29, 1494 (1984).

\bibitem{yun2} S. K. Yun, Phys. Rev. D 30, 1598 (1984).

\bibitem{chkareuli2} J. L. Chkareuli, Phys. Lett. B 300, 361 (1993).

\bibitem{barr} S. M. Barr,  Phys. Rev. D 78, 075001 (2008).

\bibitem{martinez} R. Martinez, F. Ochoa, and P. Fonseca, ``A 3-3-1 model with
$SU(8)$ unification'', arXiv:1105.4623.

\bibitem{cremmer} E. Cremmer and B. Julia, Phys. Lett. B 80, 48 (1978).

\bibitem{nielsen} N. K. Nielsen and H. R\"omer, Phys. Lett. B 154, 141 (1985).

\bibitem{duff}  M. J. Duff, ``Ultraviolet Divergences in Extended Supergravity'', in {\it Supergravity 81},
S. Ferrara and J.G. Taylor, eds., Cambridge (1982), also arXiv:1201.0386.  For the spin $\frac{3}{2}$
anomaly, see Eq. (4.6) and for the spin $\frac{3}{2}$  beta function, see Eq. (3.48).

\bibitem{marcus}  N. Marcus, Phys. Lett. B 157, 383 (1985).

\bibitem{colewein} S. Coleman and E. Weinberg, Phys. Rev. D 7, 1888 (1973).

\bibitem{georgi}  H.  Georgi and S. L. Glashow, Phys. Rev. Lett. 32, 438 (1974).

\bibitem{flipped} S. M. Barr, Phys. Lett. B 112, 219 (1982); J.-P. Derendinger, J. E. Kim, and D. V. Nanopoulos, Phys. Lett.
B 139, 170 (1984); I. Antoniadis, J. Ellis, J. S. Hagelin, and D. V. Nanopoulos, Phys. Lett. B 194, 231 (1987).  There are many
more recent papers on flipped $SU(5)$, and two that we found helpful are
A.-C. Davis and N. F. Lepora, Phys. Rev. D52, 7265 (1995), and    S. M. Barr, ref. \cite{barr1} below.

\bibitem{wiki}  Wikipedia article on ``Flipped $SU(5)$'', whose notation we follow here.   This article includes the $Z_5$ factor,
which does not appear in the original flipped $SU(5)$ articles \cite{flipped}.  It is not clear whether this was intended to indicate
a ground state modulo 5, or was just a shorthand for the division by 5 in the flipped $SU(5)$ hypercharge formula.

\bibitem{barr1}  S. M. Barr, ``Rotating Away Proton Decay in Flipped Unification'', arXiv:1307.5770.

\bibitem{curt3}  T. L. Curtright, Phys. Lett. B 102, 17 (1981).

\bibitem{fradkin}  E. S. Fradkin and A. A. Tseytlin, Phys. Lett. B 134, 301 (1984), used instanton
counting to reproduce the spin $\frac{3}{2}$
beta function coefficient given  by Curtright \cite{curt3}. I wish to
thank Arkady Tseytlin for email correspondence about this calculation.

\bibitem{peskin}  M. Peskin, ``Chiral Symmetry and Chiral Symmetry Breaking'', in
 {\it  Les Houches Summer School in Theoretical Physics: Recent Advances in Field Theory and Statistical Mechanics}, J. B. Zuber and R. Stora, eds.,   North Holland (1984).

\bibitem{thooft}  G 't Hooft, Phys. Rev. Lett. 37, 8 (1976) and Phys. Rev. D 14, 3432 (1976).

\bibitem{coleman}   S. Coleman, {\it Aspects of Symmetry}, Chapter 7 ``The uses of instantons'',
Cambridge (1985).

\bibitem{farhi} E. Farhi and L. Susskind, {\it Phys. Rep.} 74, 279 (1981).

\bibitem{mactheory}  J. M. Cornwall, Phys. Rev. D 10, 500 (1974).

\bibitem{raby}  S. Raby, S. Dimopoulos, and L. Susskind, Nucl. Phys. B 169, 373 (1980).

\bibitem{li}  L.-F. Li, Phys. Rev. D 9, 1723 (1974).


\bibitem{freedman} D. Z. Freedman and A. Van Proeyen, {\it Supergravity}, Cambridge (2012).

\bibitem{alvarez}  L. Alvarez-Gaum\'e and E. Witten, Nucl. Phys. B234, 269 (1983)

\bibitem{swein} S. Weinberg, Phys. Rev. Lett.  29, 388 (1972).

\bibitem{vacuum} S. Weinberg, {\it The Quantum Theory of Fields, Volume II: Modern Applications} (Cambridge University Press, 1996), Sec. 19.3.

\bibitem{cornwall}  J. M. Cornwall, R. Jackiw, and E. Tomboulis, Phys. Rev. D 10, 2428 (1974).

\end{thebibliography}
\end{document}